# Machine learning accelerated prediction of Ce-based ternary compounds involving antagonistic pairs

Weiyi Xia[1,2], Wei-Shen Tee[2,1], Paul C. Canfield[1,2], Fernando Assis Garcia[1,2], Raquel A Ribeiro[1,2], Yongbin Lee[1,2], Liqin Ke[1,2], Rebecca Flint[1,2], and Cai-Zhuang Wang[1,2,*]

[1]*Ames National Laboratory, U.S. Department of Energy, Iowa State University, Ames, Iowa 50011, USA*

[2]*Department of Physics and Astronomy, Iowa State University, Ames, Iowa 50011, USA*

* wangcz@ameslab.gov

## Abstract

The discovery of novel quantum materials within ternary phase spaces containing antagonistic pair such as Fe with Bi, Pb, In, and Ag, presents significant challenges yet holds great potential. In this work, we investigate the stabilization of these immiscible pairs through the integration of Cerium (Ce), an abundant rare-earth and cost-effective element. By employing a machine learning (ML)-guided framework, particularly crystal graph convolutional neural networks (CGCNN), combined with first-principles calculations, we efficiently explore the composition/structure space and predict 9 stable and 37 metastable Ce-Fe-X (X=Bi, Pb, In and Ag) ternary compounds. Our findings include the identification of multiple new stable and metastable phases, which are evaluated for their structural and energetic properties. These discoveries not only contribute to the advancement of quantum materials but also offer viable alternatives to critical rare earth elements, underscoring the importance of Ce-based intermetallic compounds in technological applications.

## 1. Introduction

An antagonistic pair, or immiscible pair, is associated with having immiscibility over almost the whole composition range, under a reasonable temperature (e.g., melting temperature) [1]. Regions of ternary phase space involving two immiscible elements, such as Fe with Pb, Bi, Ag, In, etc., remain relatively unexplored but hold great promise for novel quantum materials discovery. Ternary intermetallic compounds containing immiscible pairs are rare, yet when they do form, the immiscible elements typically segregate, with a third element encapsulating or separating them [1]. This often results in reduced dimensionality, leading to unique one-dimensional (1D) or two-dimensional (2D) structures. Forcing a 3d-transition metal (TM) like Fe to adopt such reduced dimensionality can induce complex electronic and magnetic states, including superconductivity [2], fragile magnetism [3], antiferromagnetic [4] or even ferromagnetism [1]. Especially, a recent experimental discovery of $La_4Co_4Pb$ shows distinct substructures involving antagonistic pairs, where the Co atoms adopt a corrugated Kagome net that supports itinerant antiferromagnetism [4].

On the other hand, ternary intermetallic compounds containing both Ce and Fe elements have attracted a lot of research and technological interests. Currently, high-performance magnets, essential in energy generation, conversion, and information storage devices, rely on critical RE elements such as Nd, Sm, and Dy. The insecure supply and high costs of these elements have spurred significant interest in finding alternatives. Ce, being more abundant and cost-effective, has been a viable substitute for the critical RE elements in various technological applications, particularly in clean energy and high-performance magnets [5-10]. Notably, recent studies have shown that replacing Sm with Ce and partially substituting Co with a non-magnetic element Cu can yield $CeCo_{5-x}Cu_x$ alloy with desirable magnetic properties for permanent magnet applications [11-17]. Fe being considered is also due to its abundance, cost effectiveness, and its high magnetic polarizations. These attractive features make Fe attractive for high-performance magnets development.

An interesting question is whether ternary compounds containing immiscible pairs involving Fe can be stabilized by adding Ce as the third element. To address this, a thorough understanding of the relationship between chemical compositions, crystal structures, and their relative thermodynamic stability is crucial.

We hypothesize that Ce can serve as the stabilizing third element in ternary systems involving antagonistic pairs with Fe. However, searching for stable ternary compounds involving antagonistic pairs is very challenging, and existing databases (such as the Materials Project [18], Graph Networks for Materials Exploration Database (GNoME) [19]) show that very few stable phases involving Fe-Bi, Fe-Pb, Fe-In, or Fe-Ag antagonistic pairs have been discovered. While these databases provide extensive data on material phases, it does not represent the entire spectrum of possible synthesizable materials. The limited presence of stable Fe-Bi, Fe-Pb, Fe-In, and Fe-Ag compounds may reflect an underexplored area which is promising for novel materials discovery. By integrating Ce into the Fe-X system (where X= Bi, Pb, In, Ag, is an immiscible element with Fe), we aim to explore new stable and metastable phases that leverage the unique properties of immiscible pairs within cerium-based intermetallic systems. Rapid advances in AI/ML algorithms offer great opportunities to develop new transformative strategies, and recent machine learning (ML) techniques have been employed to assist in accelerating materials design and discovery [19-30]. Our approach leverages a ML-guided framework [30] integrated with first-principles calculations to explore and identify promising candidates within the Ce-Fe-X ternary system. Specifically, we employ advanced ML techniques such as crystal graph convolutional neural networks (CGCNN) [31] to efficiently screen and predict promising candidates, thereby accelerating the materials discovery process. It should be noted that such a CGCNN approach is limited to known structure types in the database. Stable structures whose structure motifs are not presented in the existing structural databases will be missed, as shown by a recently discovered $La_4Co_4Pb$ structure by experiment [4].

The rest of the paper is organized as follows. In Section 2, we describe the machine learning (ML) methodologies employed to accelerate the discovery of new compounds, including the integration of ML with first-principles calculations and the specific algorithms used. Section 3 presents the results of our first-principles calculations guided by ML, detailing the structural and energetic properties of the identified stable and metastable compounds. Finally, in Section 4, we provide a summary of our findings, discuss the implications for future research, and suggest potential applications of the discovered compounds.

## 2. ML and Computational Methods

The workflow for our ML-guide computational approach for predicting the low-energy Ce-Fe-X (X= Bi, Pb, In, Ag) ternary compounds is illustrated in **Fig. 1**.

By extracting 28469 crystalline ternary structures from the Materials Project (MP) database [18], we substitute the three elements in these known ternary compounds with Ce, Fe, and X respectively. The template structures from MP cover most of the currently known structure motifs. No discrimination in selecting the structure motifs when these template structures are extracted from MP database to ensure unbiased ML screening to identify low-energy structures for the systems of our interest. For each ternary template structure from MP, we generate 30 structures by shuffling the order of the three elements (6 ways) and by uniformly expanding or contracting the volume of the structure by 5 scaling factors (0.92, 0.96, 1.0, 1.04, and 1.08). Varying the volume of the template structure is based on the consideration that the sizes of the elements in the template structure would be very different from those in our Ce-Fe-X systems, thus the corresponding equilibrium volumes would be very different. Varying the volume of the template structure would help in capturing better candidates by the consequent ML screening for the systems of our interest. In this way, a structure pool of 854070 hypothetical ternary compounds covering a wide range of compositions is generated for each Ce-Fe-X system. No structures relaxations are performed for these hypothetical structures before the ML screening.

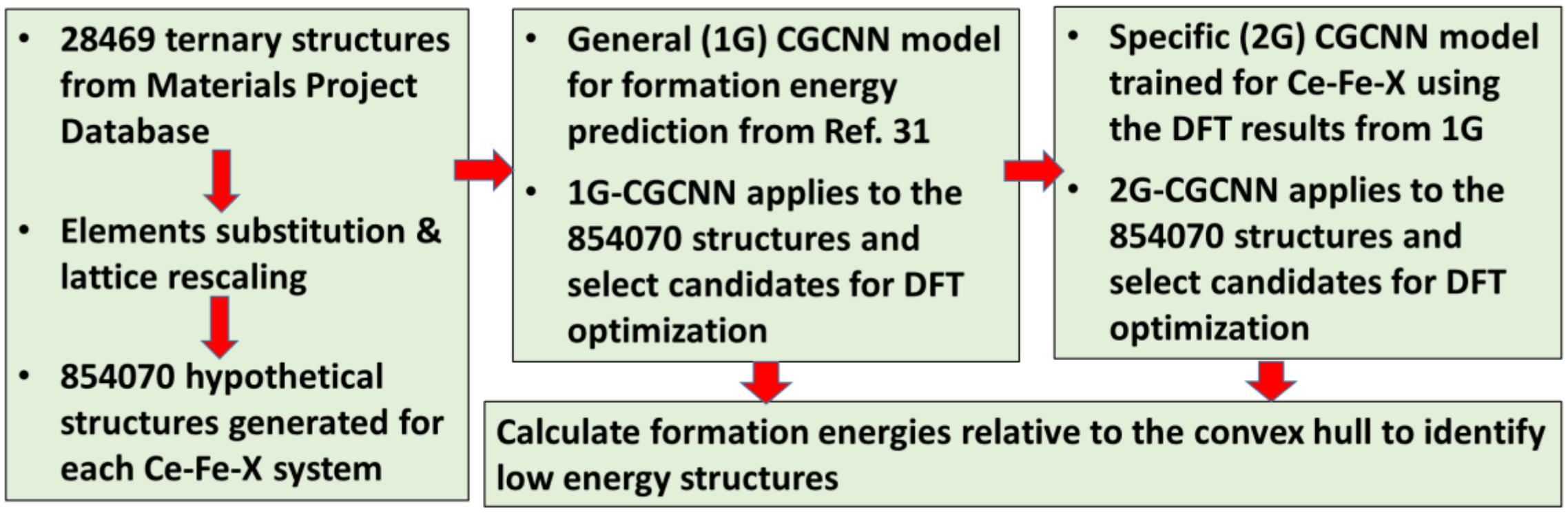

**Fig. 1.** Workflow of the ML-guided framework for predicting stable and metastable Ce-Fe-X (X = Bi, Pb, In, Ag) ternary compounds. The 1G model provides preliminary predictions based on general structural data, while the 2G model, retrained using the data from Ce-Fe-X compounds, refines these predictions with a specialized focus on the target system.

We then apply a crystal graph convolutional neural network (CGCNN) ML model to quickly evaluate the formation energies of these hypothetical ternary compounds. In CGCNN, each crystal structure is represented by a crystal graph which encodes both the atomic information and the bonding interactions between atoms in the crystal structure. The CGCNN model is then trained to predict the properties of the target crystal structures. It has been demonstrated that CGCNN ML models can be trained to efficiently evaluate the formation energies ($E_f$, see the definition below) of compounds [31]. We note that CGCNN is selected as the ML in our study due to its demonstrated ability to predict formation energies of inorganic compounds. While newer models such as ALIGNN [32] or E3NN [33] offer advantages in certain applications, the balance of simplicity, interpretability, and efficiency in CGCNN makes it well-suited for this study. Moreover, given that our approach relies on an iterative model refinement (1G to 2G discussed below), the benefits of alternate architectures are less pronounced.

In this study, we initially adopted the CGCNN model for compound formation energy prediction developed in Ref [31]. This model is trained using the structures and energies of 28046 ternary and binary compounds from first-principles density functional theory (DFT) calculations as documented in the MP database [18]. We refer to this model as 1G-CGCNN model. The training data for the 1G-CGCNN model cover many types of structures and chemical elements. This model can be used as a universal model for predict the formation energies of compounds. By applying the 1G-CGCNN ML model to screen the formation energy of the 854070 hypothetical ternary compounds, several thousand compounds with lower formation energies ($E_f \leq 0$ eV/atom) for each Ce-Fe-X system are selected for further relaxation and evaluation by first-principles calculations.

When new crystal structures and their formation energies of the Ce-Fe-X ternary compounds become available from first-principles calculations based on the candidates selected from the 1G-CGCNN screening, we train a second-generation CGCNN model (2G-CGCNN) using these system-specific data. As will be shown in the Results section, 287, 303, 541, 435 new structures for Ce-Fe-Bi, Ce-Fe-Pb, Ce-Fe-In, and Ce-Fe-Ag ternary compounds respectively are obtained from first-principles calculations based on the selections of 1G-CGCNN. The crystal structures and the formation energy $E_f$ of these 1566 new ternary compounds are used to train the 2G-CGCNN specifically for Ce-Fe-X systems. We adopt the hyperparameters for the training the same as those used to train the 1G-CGCNN model [31]. The data set of the 1566 structures is randomly divided into a training set (80%), a validation set (10%) and a test set (10%). Batch size is set to be 256 and total number of epochs is 200. Stochastic Gradient Descent is used as an optimization algorithm. The model with smallest mean-absolute-error (MAE) for the validation set from the 200 epochs is selected as the best CGCNN model from the training. Since the 2G-CGCNN model is trained using the data only from Ce-Fe-X ternary compounds, it can be regarded as a specific model for Ce-Fe-X systems. The 2G-CGCNN model is then also applied to screening the $E_f$ of the 854130 ternary structures for each Ce-Fe-X system to select additional candidates for first-principles calculations.

The first-principles calculations for the Ce-Fe-X system are performed based on density functional theory (DFT) using the VASP package [34-35]. Perdew-Burke-Ernzerhof (PBE) functional [36] combined with the projector-augmented wave (PAW) method [37] and a cutoff energy of 520 eV are used. We use a k-point grid with a mesh size of $2\pi\times0.025$ Å$^{-1}$ generated by the Monkhorst-Pack scheme. This mesh size is fine enough to sample the first Brillouin zone for achieving better k-point convergence [38]. The lattice vectors and the atomic positions of candidate structures selected from 1G- and 2G-CGCNN predictions are fully optimized by the DFT calculations until forces on each atom are less than 0.01 eV/atom.

The formation energy $E_f$ per atom is defined relative to the elemental phases of a $\mathrm{Ce}_\alpha\mathrm{Fe}_\beta\mathrm{X}_\gamma$ with $\alpha+\beta+\gamma=1$ as

$$E_f = E\left(\mathrm{Ce}_\alpha\mathrm{Fe}_\beta\mathrm{X}_\gamma\right) - \alpha E(\mathrm{Ce}) - \beta E(\mathrm{Fe}) - \gamma E(\mathrm{X}).$$

Here, $E\left(\mathrm{Ce}_\alpha\mathrm{Fe}_\beta\mathrm{X}_\gamma\right)$ is the total energy per atom of a $\mathrm{Ce}_\alpha\mathrm{Fe}_\beta\mathrm{X}_\gamma$ structure. Reference energies are the total energies per atom of face-centered cubic Ce, bcc Fe, and most stable elementary phases of X (X=Pb, Bi, Ag, In).

While the formation energies $E_f$ are used for ML screening because these target values are better learned and predicted by CGCNN, the thermodynamic stability of a ternary compound is determined by formation energy of the compound relative to the convex hull of the ternary system. For all the ternary compounds obtained after the DFT calculations, we also calculate their formation energy relative to the convex hull, $E_{\mathrm{hull}}$, by DFT. This is done by comparing the formation energy of $\mathrm{Ce}_\alpha\mathrm{Fe}_\beta\mathrm{X}_\gamma$ with respect to the nearby three known stable phases. The chemical compositions of these phases are located at the vertexes of the Gibbs triangle that encloses the composition of $\mathrm{Ce}_\alpha\mathrm{Fe}_\beta\mathrm{X}_\gamma$. We use this construction to assess the thermodynamic stability against decomposition into the stable phases. The $E_{\mathrm{hull}}$ is the decomposition energy of a $\mathrm{Ce}_\alpha\mathrm{Fe}_\beta\mathrm{X}_\gamma$ ternary compound with respect to the nearby three known stable phases which can be ternary, binary, or elemental phases.

In this study, a stable structure at T=0K is defined as it has the formation energy on the convex hull ($E_{hull}$ = 0 eV/atom), while a metastable structure is defined as it has a $E_{hull}$ ≤ 0.1 eV/atom above the convex hull. It has been shown by DFT calculations that about 50% of experimental synthesized structures in the Inorganic Crystal Structure Database have $E_{hull}$ values above the convex hull and some structures can have $E_{hull}$ as high as 0.1 eV/atom [38]. Therefore, predicted structures with $E_{hull}$ below 0.1 eV/atom can be considered as experimentally synthesizable structures under non-equilibrium synthesis conditions.

## 3. Results

The formation energy distributions of hypothetical Ce-Fe-X compounds, as predicted by 1G- and 2G-CGCNN models, are shown in **Fig. 2**. The 1G model, trained on a diverse set of binary and ternary compounds from the Materials Project database, provides an initial screening of low-energy candidates for our Ce-Fe-X systems. The 2G model, trained specifically on Ce-Fe-X compounds, refines the predictions by leveraging system-specific data. It can be seen from **Fig. 2** that the formation energy distribution predicted by the 2G-CGCNN model is systematically shifted to higher energy as compared to the distribution from the 1G-CGCNN model.

After eliminating redundant structures, we select approximately 2,000 candidates with lower formation energies ($E_f$ ≤ 0.0 eV/atom) from the 1G-CGCNN predictions and 300-600 structures (i.e., 330, 567, 617 and 233 structures for X=Bi, Pb, In, Ag, respectively) from the 2G-CGCNN predictions for first-principles calculations. It should be noted that about 96% of the structures selected by the 1G-CGCNN which have the formation energy within 100 meV/atom above the convex hull by DFT calculations have also been recaptured by the 2G-CGCNN. These duplicated structures from both 1G and 2G are not included in the numbers listed above from the 2G-CGCNN selection to avoid duplicating DFT calculations. Therefore, the predictions from 2G-CGCNN help to capture more stable and metastable Ce-Fe-X compounds.

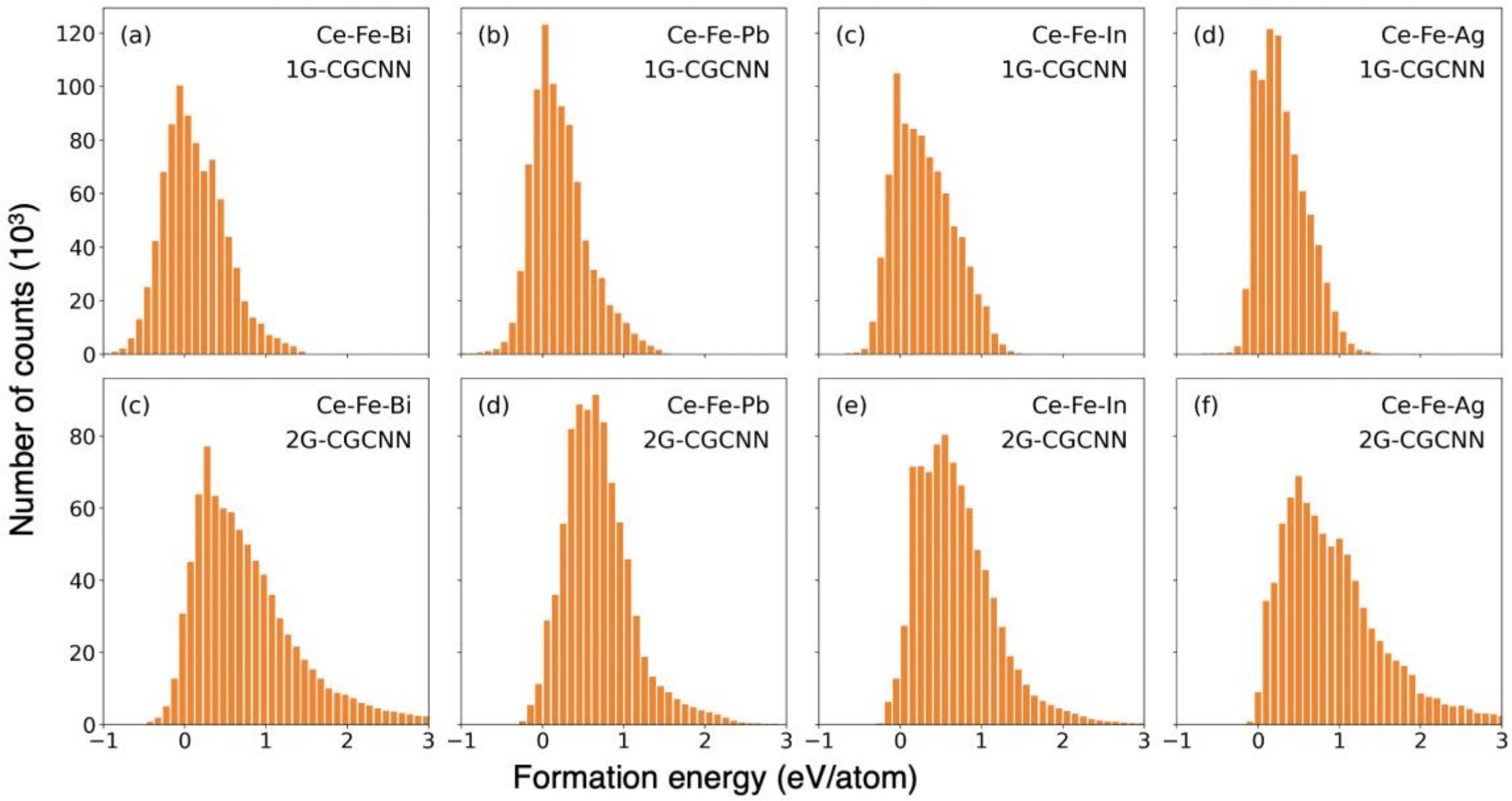

**Fig. 2.** Distribution of formation energies ($E_f$) of the hypothetical ternary Ce-Fe-X ternary compounds (X=Bi, Pb, In, Ag from left to right) predicted from the 1G (top), and 2G (bottom) CGCNN energy models. The total number of structures for each Ce-Fe-X system is 854,070. The 1G model provides preliminary predictions based on general structural templates, while the 2G model, retrained on Ce-Fe-X compounds, refines these predictions with a specialized focus on the target system

The results from the DFT calculations show that 287 non-equivalent Ce-Fe-Bi structures selected from 1G- and 111 from 2G-CGCNN predictions can be fully optimized. Other structures that cannot pass the electronic self-consistent calculations are discarded. These discarded structures are most likely to be far from the realistic structures for Ce-Fe-Bi ternary compounds. For Ce-Fe-X (X=Pb, In, Ag), there are 303, 541, 435 structures from 1G-, and 164, 157, 57 from 2G-CGCNN predictions, respectively. Notably, the structures from 2G-CGCNN selection have a higher success rate for DFT relaxation and are more energetically favorable than those from 1G-CGCNN selection.

The different performances between the 1G- and 2G-CGCNN models can be understood by comparing the formation energies ($E_f$) predicted by the CGCNN models to those from DFT calculations for the Ce-Fe-X ternary compounds. Two sets of structures are used for these comparisons. The first set is the1566 Ce-Fe-X (X=Bi, Pb, In, Ag) ternary structures from the DFT relaxations based on the candidates selected from the 1G-CGCNN. The comparisons are shown in **Fig. 3 (a)** and **(b)** respectively. In comparison with the DFT results, the 1G-CGCNN model trends overestimate the formation energies for the Ce-Fe-X ternary compounds with most of the points well below the diagonal line in **Fig 3 (a)** and with a large systematic mean absolute error (MAE) ~ 0.258 eV/atom presumably due to the use of data covering a wide range of chemical elements in the training. By contrast, the formation energy from the 2G-CGCNN model agrees much better with the DFT results (MAE ~0.046 eV/atom), indicating the 2G-CGCNN model is accurately trained for Ce-Fe-X systems. This comparison explains why the formation energy distribution predicted by the 1G-CGCNN model is systematically lower than that predicted by the 2G-CGCNN model as shown in **Fig. 2**. Another set of structures used for the comparison is the 665 structures relaxed by DFT based on the 2G-CGCNN selection. These structures are neither used in 1G nor 2G-CGCNN training, so they can serve as the testing set for the prediction accuracy of the CGCNN models. The formation energies predicted from 1-G and 2G-CGCNN models for this testing set are compared with the results from DFT calculations as shown and in **Fig. 3 (c)** and **(d)** respectively. We can see from **Fig. 3 (c)** that the 1G-CGCNN model also systematically overestimates the formation energies for this testing set with a MAE ~ 0.157 eV/atom. In comparison, the 2G-CGCNN model achieves a MAE of 0.045 eV/atom in formation energy prediction, a significant improvement over the 1G-CGCNN model. Nevertheless, as one can see from **Fig. 3 (a)** and **(c)**, the 1G-CGCNN predicts the correct trend of the formation energy for Ce-Fe-X ternary compounds albeit with large error bars. It is a reasonable choice to use 1G-CGCNN model to initially screen the relative stability of new ternary compounds for any system of interest. The results from this universal 1G-CGCNN can be used to train the consequent more accurate CGCNN models for specific system of interest. Such an iterative improvement of CGCNN model can be repeated until satisfactory accuracy is reached.

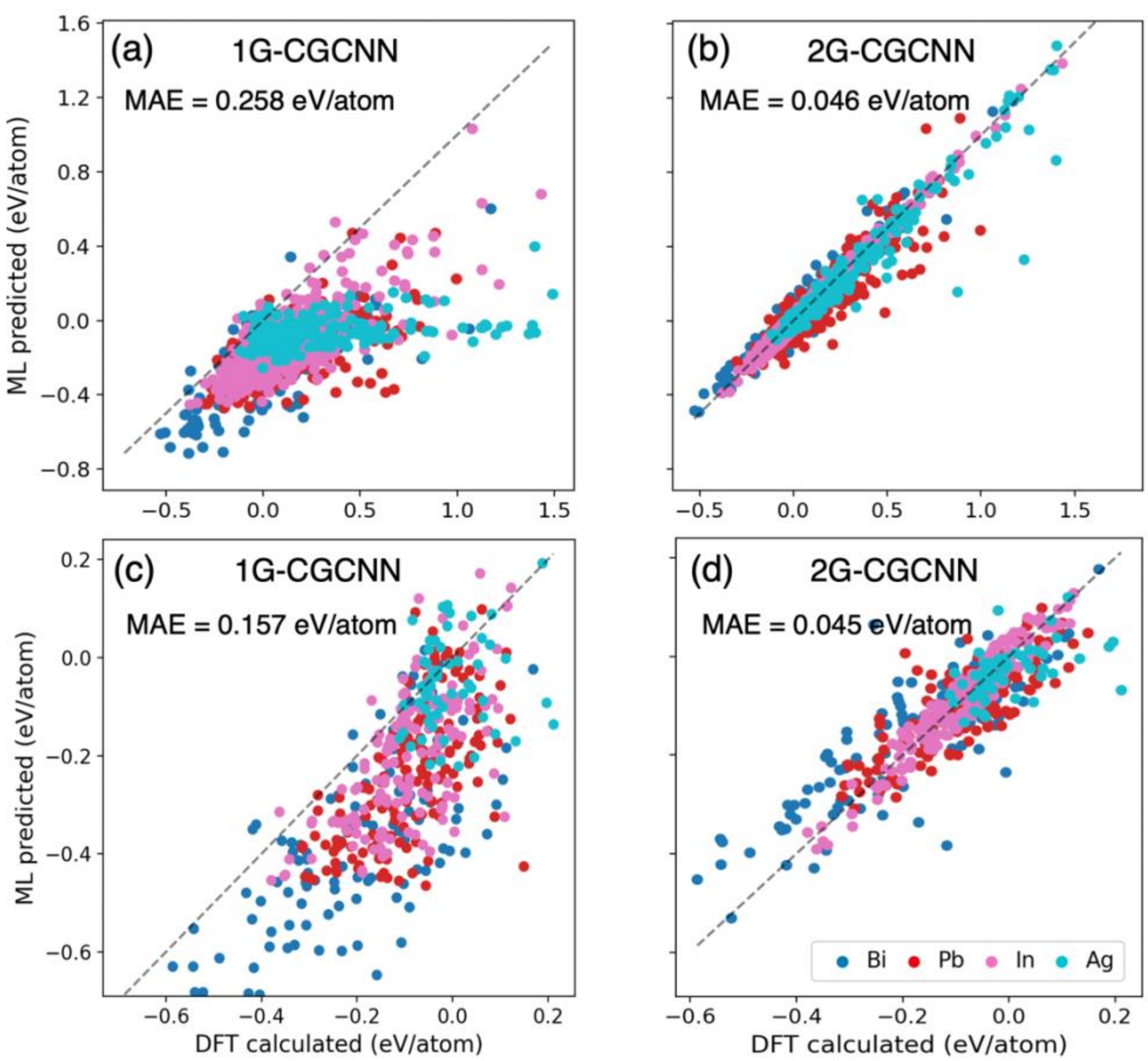

**Fig. 3.** Comparison the accuracy of the 1G- and 2G-CGCNN model for formation energy predictions of Ce-Fe-X ternary compound with respect to the results from DFT calculations. The 1566 Ce-Fe-X ternary structures from 1G-CGCNN selection and subsequent DFT calculations are used in (a) and (b); The 1566 Ce-Fe-X ternary structures from 1G-CGCNN selection and subsequent DFT calculations are used in (a) and (b); The 665 Ce-Fe-X ternary structures from 2G-CGCNN selection and subsequent DFT calculations are used in (c) and (d). Different colors correspond to different element X (Bi, Pb, In and Ag).

To evaluate the thermodynamic stability of these newly predicted ternary compounds, we then calculate the formation energies ($E_{hull}$) of these structures with respect to the Ce-Fe-X ternary convex hull at the accuracy level of DFT. The compositions of stable (with $E_{hull} \leq 0$ eV/atom) and low-energy metastable (with $E_{hull} \leq 0.1$ eV/atom) Ce-Fe-X ternary phases with respect to the currently known convex hull predicted from our CGCNN+DFT calculations can be seen from **Fig. 4** where all predicted ternary Ce-Fe-X compounds with $E_{hull}$ below 0.1 eV/atoms are shown. More detailed information about the compositions and structure symmetries of those ternary compounds with $E_{hull}$ on the convex hull or within 50 meV/atom above the convex hull is shown in **Table 1**. The predicted ternary Ce-Fe-X compounds with $E_{hull}$ between 50-100 meV/atom are given in the Supplementary Materials.

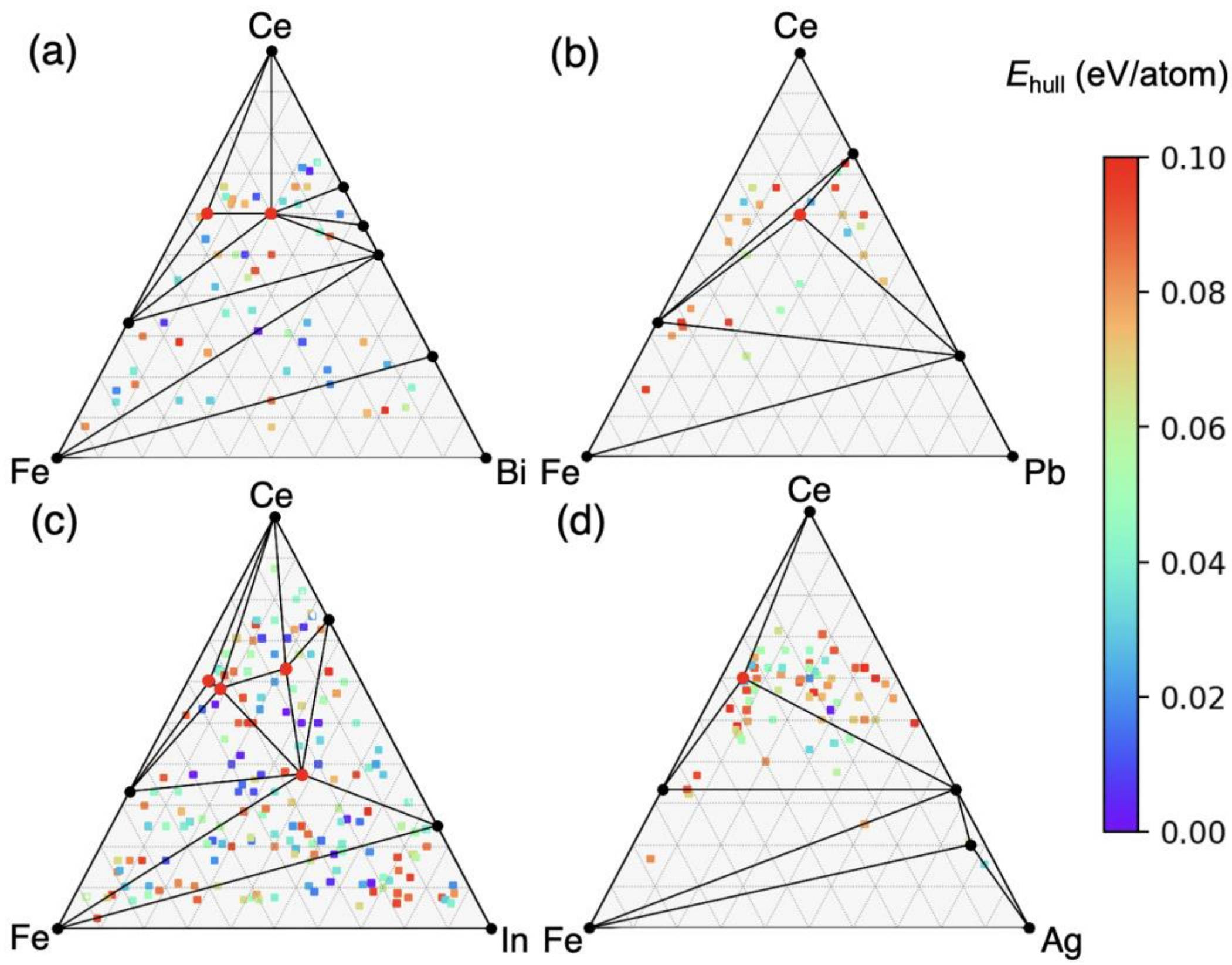

**Fig. 4.** The updated convex hull covering a wide range of compositions and with $E_{hull} \leq 0.1$ eV/atom for (a) Ce-Fe-Bi (b) Ce-Fe-Pb (c) Ce-Fe-In (d) Ce-Fe-Ag ternary phases predicted from our CGCNN+DFT approach. The black dots on the vertices of the convex hull are existing stable phases and the red dots on the vertices are newly predicted stable

phases. The compositions in the convex hull are colored according to the lowest- $E_{\text{hull}}$ for this composition. The $E_{\text{hull}}$ shown in the color bars are in the unit of eV/atom.

From the results shown in **Fig. 4** and **Table 1**, we can see that the CGCNN search covers a wide range of compositions and a noticeable number of structures are predicted to have formation energies either under the convex hull or within 50 meV/atom above the convex hull (dark blue color) based on currently known stable binary and elemental phases. We also list all other structures with formation energies less than 100 meV/atom above the convex hull (light blue) in Supplemental Materials (**Table S1**). There are two new stable ternary phases predicted for Ce-Fe-Bi system, one for Ce-Fe-Pb, four for Ce-Fe-In and two for Ce-Fe-Ag, respectively. We note that the stability presented here is based on DFT energies at T= 0 K. No partial site occupancies are considered. Allowing partial site occupancies may further lower the free energies at finite temperatures due to the entropy contribution. We also note that many stable and metastable compounds (as shown in dark blue and purple in **Fig. 4**) obtained from our predictions are clustering at Ce-rich regions in the ternary convex hull. This result indicates that these predicted stable phases may compete for phase selection and stability in synthesis. Interestingly, there are some structures predicted to be metastable in the Fe-rich region, which would be potential candidates for magnetic materials.

**Structures of Ce-Fe-Bi compounds -** For Ce-Fe-Bi system, we predicted 2 stable phases, along with 6 metastable phases with $E_{\text{hull}} \leq 50$ meV/atom. In **Fig. 5**, we show the 2 stable compounds (**Fig. 5 (a)** – **5 (b)**) and 2 lowest-energy metastable structures, with $E_{\text{hull}} = 4$, 18 meV/atom (**Fig. 5 (c)** – **5 (d)**), respectively. Detailed structural information of all stable and metastable phases are given in Table 1.

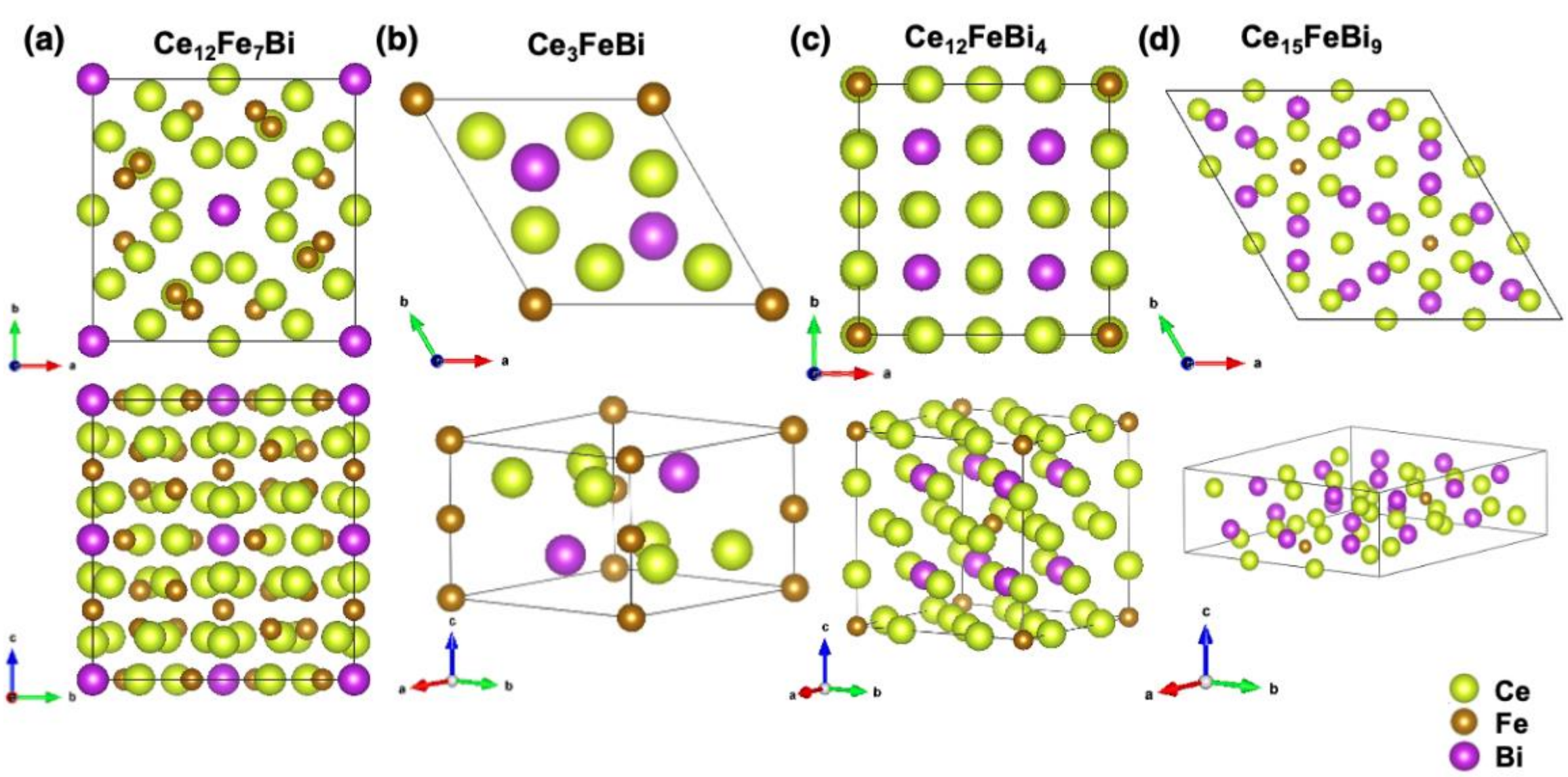

**Fig. 5.** The structures of 2 stable (a-b) and 2 metastable (c-d) Ce-Fe-Bi structures obtained from our predictions. More detailed information about these structures is also given in **Table 1**.

The predicted stable $Ce_{12}Fe_7Bi$ compound has a tetragonal lattice with *I4/mcm* space group, as plotted in **Fig. 54** (a). There are three Wyckoff sites for Ce and Fe, one Wyckoff site for Bi. The Ce-Fe bond lengths range from 2.62 to 3.12 Å, and the Ce-Bi bond length is 3.28 Å. There is one Fe-Bi bond (along out-of-plane direction) with length of 3.13 Å, forming a 1D chain along c direction, indicating the immiscible pair Fe-Bi is not well separated in this structure.

Another predicted stable $Ce_3FeBi$ compound crystallizes in a hexagonal hcp lattice, as plotted in Fig. **5 (b)**. Ce is bonded to two equivalent Fe atoms and four equivalent Bi atoms. The Ce-Fe bond length is 2.63 Å, while there are one shorter Ce-Bi bonds with 3.43 Å and two longer ones with 3.52 Å. The immiscible Fe-Bi pairs are weakly bonded with bond length of 4.31 Å..
The rest two compounds, $Ce_{12}FeBi_4$, $Ce_{15}FeBi_9$, are calculated to be less than 20 meV/atom above the convex hull (see Table 1). The Fe-Bi pairs are well separated in the four predicted higher-energy structures, as shown in **Fig. 5 (c)** – **5 (d)**.

**Structures of Ce-Fe-Pb compounds -** For the Ce-Fe-Pb system, we predict one stable phase, along with 5 metastable phases with $E_{\text{hull}} \leq 50$ meV/atom. In **Fig. 6**, we show the stable phase (**Fig. 6 (a)** and 3 lowest-energy metastable structures, with $E_{\text{hull}} = 23, 26, 27$ meV/atom (**Fig. 6 (b) – 65 (d)**), respectively. Detailed structural information of all stable and metastable phases are given in Table 1.

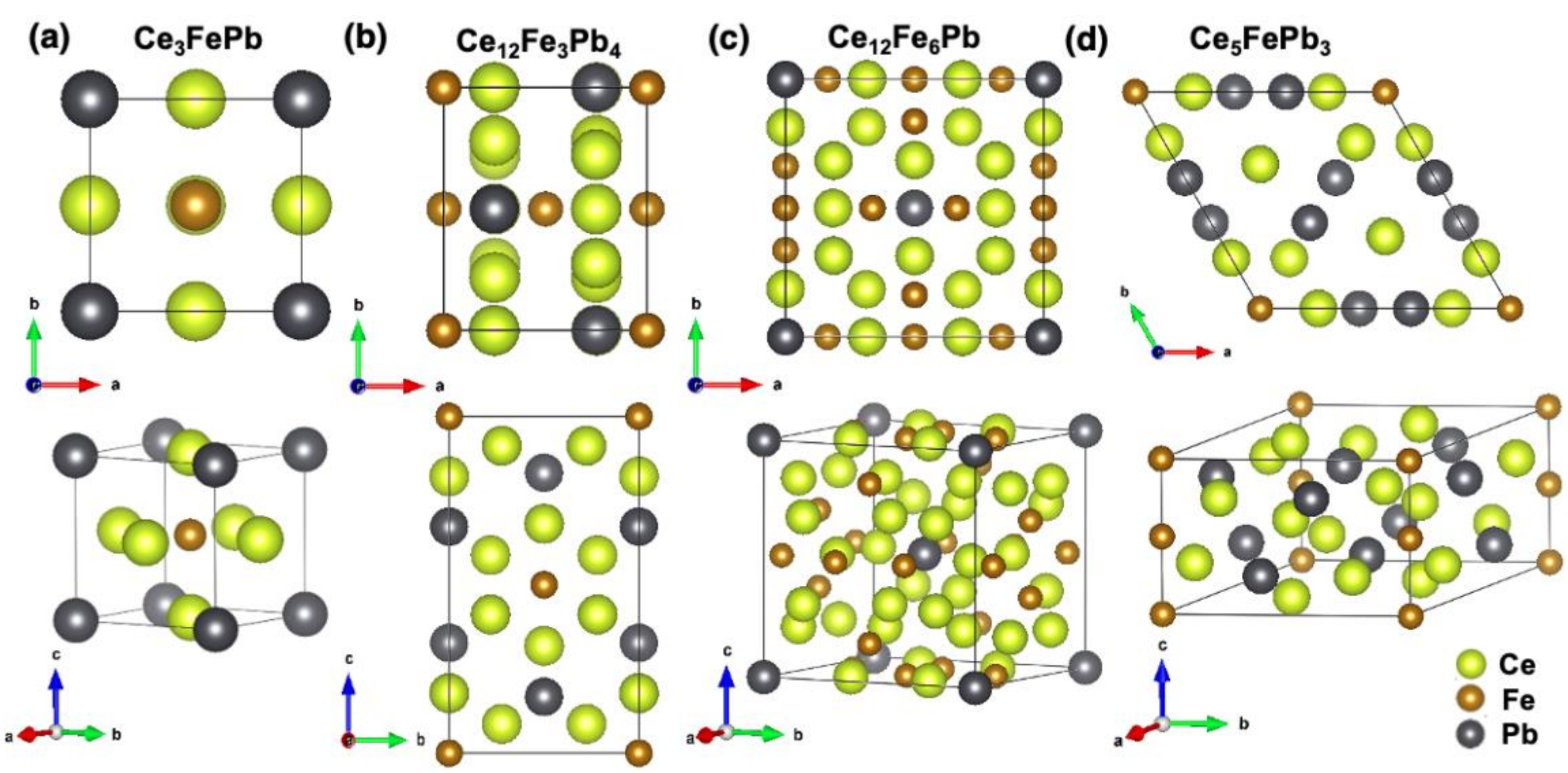

**Fig. 6.** The structures of one stable (a) and three metastable (b-d) Ce-Fe-Pb structures obtained from our predictions. More detailed information about these structures is also given in **Table 1**.

The crystal structure of the predicted stable $Ce_3FePb$ compound is a typical perovskite structure, with a *Pm-3m* space group symmetry and a lattice constant of 5.11 Å at the GGA-PBE level of calculation. There is one Wyckoff site for each of Ce, Fe and Pb. the atoms of the immiscible pair Fe-Pb are completely encapsulated by Ce atoms. This structure has also been reported by Ref [19] in GNoME database.

The other three compounds, $Ce_{12}Fe_3Pb_4$, $Ce_{12}Fe_6Pb$, $Ce_5FePb_3$, are calculated to have formation energy within 30 meV/atom above the convex hull (see **Table 1**), who's structures are shown in **Fig. 6 (b)** – **6 (d)**. We can see from the plots that in these three metastable phases, Fe and Pb atoms are completely separated.

**Structures of Ce-Fe-In compounds -** For the Ce-Fe-In system, we predicted four stable phases, along with 12 metastable phases with $E_{hull} \leq 50$ meV/atom. In **Fig. 7**, we show the 4 stable compounds. Detailed structural information of all stable and metastable phases are given in Table 1. We fail to capture another stable phase, $Ce_3Fe_2In_2$, reported by ref [19], owing to the missing structural motif in MP database. The predicted formation energy for this phase with 1G and 2G-CGCNN model is -0.25 meV/atom and -0.2 meV/atom, respectively, which indicates that this phase will be captured if the motif exists in MP database.

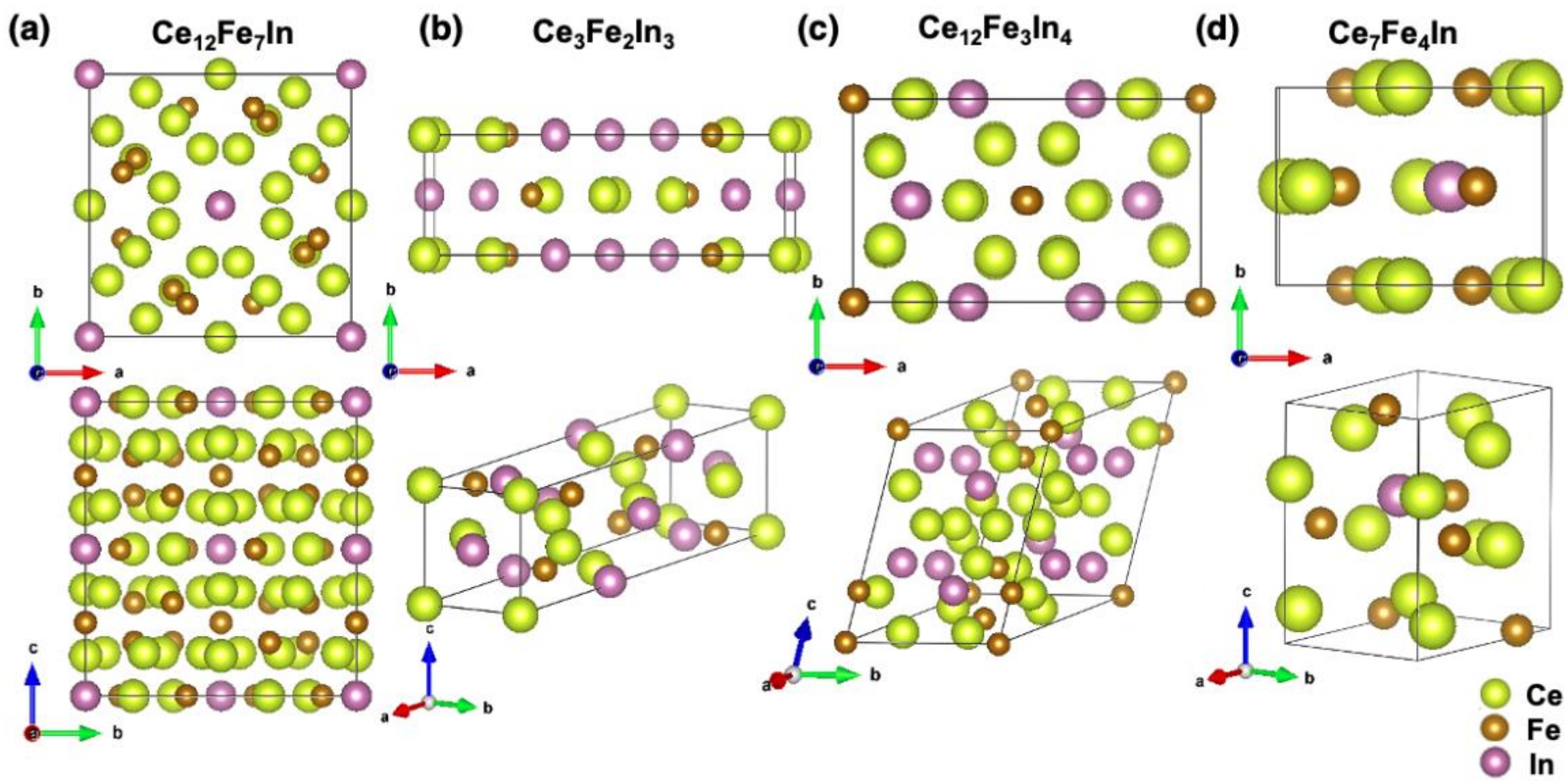

**Fig. 7.** The structures of 4 stable Ce-Fe-In structures obtained from our predictions. More detailed information about these structures is also given in **Table 1**.

The $Ce_{12}Fe_7In$ compound shown in **Fig. 7(a)** is very similar to the predicted stable $Ce_{12}Fe_7Bi$ compound. Notably, the bond length of Fe-In is 3.14 Å.

The second compound predicted to be stable is the $Ce_3Fe_2In_3$ compound, as plotted in **Fig. 7 (b)**. It is a monoclinic structure with a C2/m space group symmetry. There are two Wyckoff sites for Ce and In, one for Fe. Fe is bonded to three equivalent In atoms, with bond lengths of 2.67 Å and 2.70 Å.

The $Ce_{12}Fe_3In_4$, as plotted in **Fig. 7 (c)**, has a monoclinic lattice with a C2/m space group symmetry. There are four Wyckoff sites for Ce, two for Fe and In. The first Fe site is bonded to 6 Ce atoms forming a distorted FeCe6 octahedra that shares corners with $InCe_{12}$ cuboctahedra. Notably, it shows perfect segregation of immiscible pair Fe-In, where Fe atom is completely encapsulated by Ce atoms.

And the $Ce_7Fe_4In$ compound plotted in **Fig. 7 (d)** is a monoclinic structure with *Pm* space group symmetry. There are 7 Wyckoff sites for Ce, 4 for Fe and one for In. The In atom is bonded in a 13-coordinate geometry to 9 Ce atoms with bond length ranging from 3.23 Å to 3.73 Å, and 4 Fe atoms with bond length ranging from 2.82 Å to 2.85 Å, indicating that Fe and In pair is not separated here.

**Structures of Ce-Fe-Ag compounds -** For the Ce-Fe-Ag system, we predict two stable phases, along with 14 metastable phases with $E_{hull} \leq 50$ meV/atom. In **Fig. 8**, we show the two stable compounds (**Fig. 8 (a) – 8 (b)**) and 2 lowest-energy metastable structures, with $E_{hull}$ = 1.6, 1.7 meV/atom (**Fig. 8 (c) – 8 (d)**), respectively. Detailed structural information of all stable and metastable phases are given in Table 1.

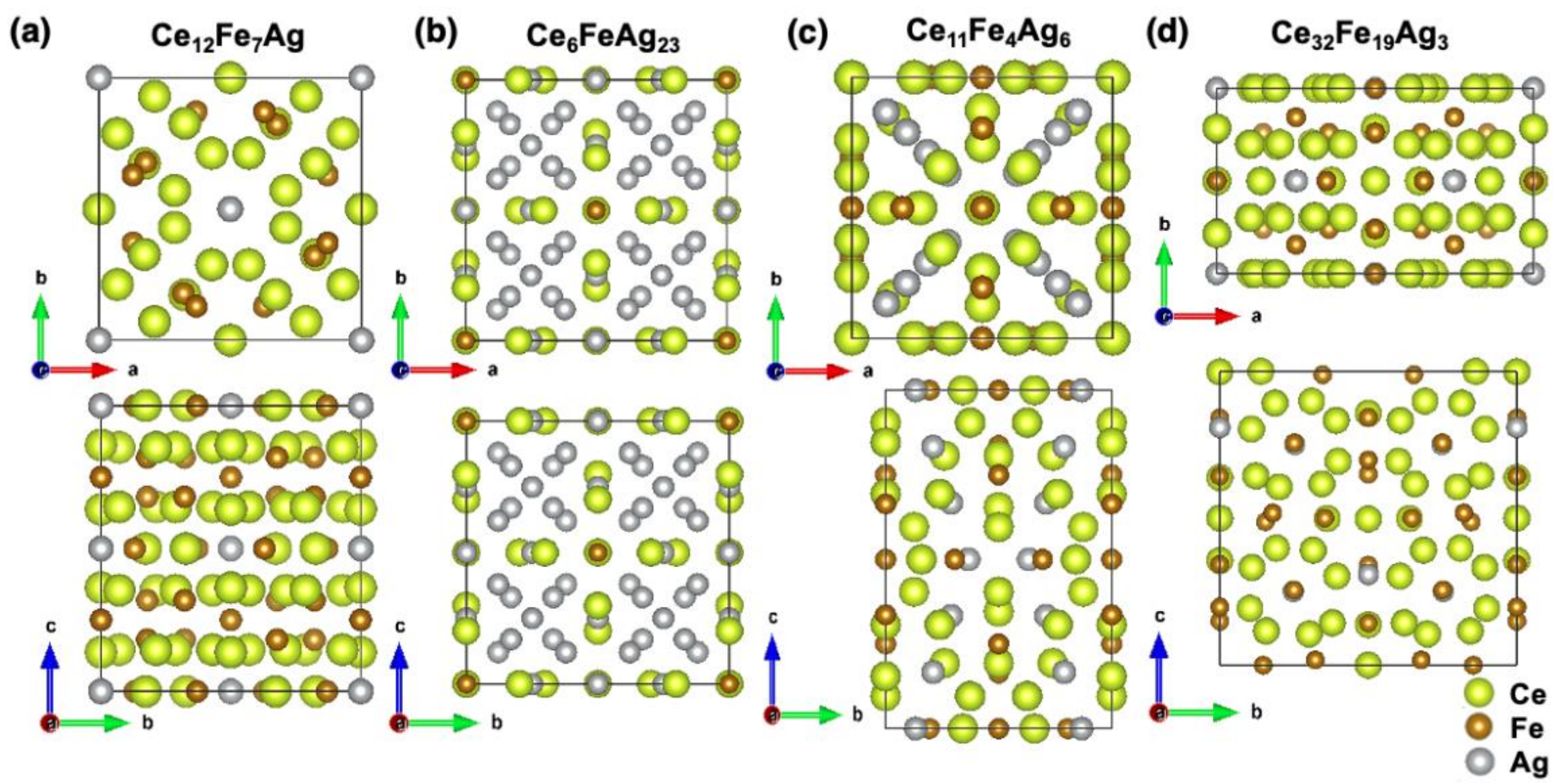

**Fig. 8.** The structures of 2 stable (a-b) and 2 metastable (c-d) Ce-Fe-Ag structures obtained from our predictions. More detailed information about these structures is also given in **Table 1**.

The predicted stable $Ce_{12}Fe_7Ag$ compound is very similar to the predicted stable $Ce_{12}Fe_7Bi$ compound. Notably, the bond length of Fe-Ag is 3.16 Å, so again the immiscible pair are not well separated.

The crystal structure of the predicted stable $Ce_6FeAg_{23}$ compound is shown in **Fig. 8 (b)**. It has a tetragonal lattice with *Fm-3m* space group symmetry. There are 4 Wyckoff sites for Ag, and one for Ce and Fe. The Fe atom is bonded to 6 Ce atoms to form $FeCe_6$ octahedra, which share edges with 12 $Ce_3Ag_9$ cuboctahedra. We can clearly see the segregation of Fe and Ag atoms.

The other two compounds, $Ce_{11}Fe_4Ag_6$, $Ce_{32}Fe_{19}Ag_3$, as shown in **Fig. 8 (c)** – **8 (d)**, are all predicted to have formation energy within 30 meV/atom above the convex hull (see Table 1). Notably, these two compounds have formation energies only 2 meV/atom above the updated convex hull. In both compounds, we observe complete segregation of the immiscible pair Fe-Ag.

Interestingly, among the predicted stable and metastable structures, there are repeated motifs for multiple systems, i.e., a 12-7-1 (i.e., $Ce_{12}Fe_7X$, same abbreviation below) motif with *I4/mcm* space group for X=Bi, In, Ag; a 12-3-4 motif for X=Bi, Pb, In; a 3-1-1 motif with *Pm-3m* space group for X=Pb, In, (For Bi and Ag, the stable 3-1-1 phases have different motifs, while the phase with the same motif has higher energy); a 12-1-4 motif with *Im-3m* space group for X=Bi and Pb; and a 12-6-1 motif for Pb and Ag. Several of the predicted stable structures show perfect segregation of immiscible pairs by the Ce atoms.

## 4. Concluding Remarks

In summary, we explore the uncharted regions of ternary phase space involving antagonistic pairs of Fe with Pb, Bi, Ag, and In, aiming to discover novel quantum materials. We focus on integrating Ce as the third element to stabilize these immiscible pairs, leveraging its relative abundance and cost-effectiveness. Utilizing a machine learning (ML)-guided framework combined with first-principles calculations, we predict various Ce-Fe-X stable and metastable ternary compounds. This approach employs a crystal graph convolutional neural network (CGCNN) to estimate the formation energies of millions of hypothetical Ce-Fe-X compounds. By efficiently screening these compounds, only a few hundred promising structures are selected for detailed DFT calculations, significantly reducing computational effort compared to traditional high-throughput methods. This strategy dramatically speeds up our search for new stable and metastable phases, including two stable phases for Ce-Fe-Bi, one for Ce-Fe-Pb, four for Ce-Fe-In, and two for Ce-Fe-Ag, along with 37 metastable phases with $E_{hull} \leq 50$ meV/atom, and more phases with $E_{hull} \leq 100$ meV/atom (listed in Supplemental Materials). These discoveries suggest promising applications in magnetic materials and highlight the potential of Ce-based intermetallic compounds as substitutes for critical rare earth elements. Thus, the ML-guided framework presented here represents a powerful new paradigm for material design and discovery, demonstrating its effectiveness in efficiently searching for stable complex compounds across a wide range of chemical and structural spaces.

Despite the success of our approach, it is essential to note that not all predicted materials will be synthesizable in real life. First, the formation energy predictions are performed at 0 K, which alone is not sufficient for determining chemical stability. Additional validation using phonon calculations and ab initio molecular dynamics (AIMD) simulations would provide further insights into dynamical stability, which will be explored in detail in our consequent studies. The results from present work provide candidates that warrant further theoretical investigation and possibly subsequent experimental synthesis and characterization. . Second, this raises an interesting open question as to *why* – what might this approach be missing about the physics of synthesizing novel materials with antagonistic pairs? Many known ternary intermetallic compounds containing immiscible pairs exhibit a structure motif in which the immiscible elements are spatially separated and encapsulated by the third element which provides the stability to the formation of the ternary compounds. This structural arrangement is also evident in the predicted Ce-Fe-X compounds presented in this paper. Among the 37 structures with $E_{hull}$ within 50 meV/atom, 26 show clear segregation of the antagonistic pair, while 11 do not. Experimentally, there are no known cases of unsegregated antagonistic pairs so far, indicating that our predictions of stable compounds with unsegregated antagonistic pairs need to be experimentally verified. If segregation is required or enforced by nature, the prediction of non-segregated compounds suggests there may be undiscovered compounds (e.g., new structure types) that could change the hull and affect stability predictions. Conversely, if a new compound/composition is found experimentally to be stable but missed by the current algorithm, this would alter the hull and predictions.

It should be noted that in this approach, stable structures whose motifs are not present in existing structural databases will be missed, as is common in most current AI/ML approaches. Integrating available crystal structure prediction methods, such as genetic algorithm (GA) [39-41], for new structure search based on the promising compositions predicted by CGCNN should help overcome this deficiency. From this perspective, predictions from the CGCNN-guided approach can efficiently locate low-energy basins in complex compositional and structural spaces. Additionally, successful synthesis of the predicted materials will require a better understanding of the local energy landscape between transient or intermediate products, temperature stability, and crystal nucleation and growth kinetics of the predicted ternary compounds and nearby competing phases. Future computational work should also address dynamical stability using phonon calculations and the kinetics of phase selection and formation using MD simulations. Moreover, a fundamental scientific question remains: "*What third element, and in what fraction, can be added to make stable ternary compounds out of immiscible pairs of elements, and why?*" These problems should be the focus of future studies.

**Acknowledgments**

Work at Ames National Laboratory was supported by the U.S. Department of Energy (DOE), Office of Science, Basic Energy Sciences, Materials Science and Engineering Division including a grant of computer time at the National Energy Research Supercomputing Center (NERSC) in Berkeley. Ames National Laboratory is operated for the U.S. DOE by Iowa State University under contract # DE-AC02-07CH11358.

**Data availability**

The data leading to the findings in this paper and the ML models are available from the authors upon reasonable request.

## References

[1] P.C. Canfield, New Materials Physics, Rep. Prog. Phys., **83**, 016501 (2020).

[2] P. C. Canfield and S. L. Bud'ko, FeAs-Based Superconductivity: A Case Study of the Effects of Transition Metal Doping on BaFe2As2, Annual Review of Condensed Matter Physics 1(1):27-50 (2010).

[3] U.S. Kaluarachchi, S.L. Bud'ko, P.C. Canfield, V. Taufour, Tricritical wings and modulated magnetic phases in $LaCrGe_3$ under pressure, Nature Communications, **8**, 546 (2017).

[4] T. J. Slade, N. Furukawa, M. Dygert, S. Mohamed, A. Das, W. Xia, C. Z. Wang, S. L. Budko, P. C. Canfield, $La_4Co_4X$ (X = Pb, Bi, Sb): a demonstration of antagonistic pairs as a route to quasi-low dimensional ternary compounds, Phys. Rev. Mater. **8**, 064401.

[5] E. A. Nesbitt, R. H. Willens, R. C. Sherwood, E. Buehler, and J. H. Wernick, New permanent magnet materials, Applied Physics Letters **12**(11), 361–362 (1968).

[6] D. Gignoux, F.Givord, R. Lemaire, H. Launois, and F. Sayetat, Valence state of cerium in the hexagonal $CeM_5$ compounds with the transition metals, Journal de Physique **43**, 173–180 (1982).

[7] E. Bauer, E. Gratz, and C. Schmitzer, CeCu5: Another Kondo lattice showing magnetic order, Journal of Magnetism and Magnetic Materials, **63-64**, 37–39 (1987).

[8] E. Bauer, Anomalous properties of Ce-Cu- and Yb-Cu-based compounds, Advances in Physics **40**(4), 417–534 (1991).

[9] P. Coleman, Handbook of Magnetism and Advanced Magnetic Materials, Vol. 1 (Wiley, New York, 2007) pp. 95–148.

[10] Q. Si and F. Steglich, Heavy Fermions and Quantum Phase Transitions, Science **329**, 1161 (2010).

[11] Y. Tawara and H. Senno, Cerium, cobalt and copper alloy as a permanent magnet material, Japanese Journal of Applied Physics **7**(8), 966–967 (1968).

[12] T. N. Lamichhane, M. T. Onyszczak, O. Palasyuk, S. Sharikadze, T. -H. Kim, Q. Lin, M. J. Kramer, R.W. McCallum, A. L. Wysocki, M. C. Nguyen, V. P. Antropov, T. Pandey, D. Parker, S. L. Bud'ko, P. C. Canfield, and A. Palasyuk, Single-crystal permanent magnets: Extraordinary magnetic behavior in the Ta-, Cu-, and Fe-substituted CeCo5 systems, Phys. Rev. Applied **11**, 014052 (2019).

[13] H. Shishido, T. Ueno, K. Saito, M. Sawada, and M. Matsumoto, Intrinsic coercivity induced by valence fluctuations in 4f -3d intermetallic magnets, arXiv:2103.10202 (2021).

[14] D. Girodin, C.H. Allibert, F. Givord, and R. Lemaire, Phase equilibria in the CeCo5-

CeCu5 system and structural characterization of the Ce(Co1−xCux)5 phases, Journal of the Less Common Metals **110**(1), 149–158 (1985).

[15] F. Meyer-Liautaud, Souad Derkaoui, C.H. Allibert, and R. Castanet, Structural and thermodynamic data on the pseudobinary phases R(Co1−xCux)5 with R = Sm, Y, Ce. Journal of the Less Common Metals **127**, 231–242 (1987).

[16] M. Matsumoto, Magnetism trends in doped Ce-Cu intermetallics in the vicinity of quantum criticality: Realistic Kondo lattice models based on dynamical mean-field theory, Phys. Rev. Materials **4**, 054401 (2020).

[17] A. M. Guloy and J. D. Corbett, Cu substituted CeCo5: New optimal permanent magnetic material with reduced criticality, Journal of Alloys and Compounds **723**, 208-212 (2017).

[18] A. Jain, S. P. Ong, G. Hautier, W. Chen, W. D. Richards, S. Dacek, S. Cholia, D. Gunter, D. Skinner, G. Ceder, K. A. Persson, Commentary: The Materials Project: A materials genome approach to accelerating materials innovation, APL Mater. **1**, 011002 (2013).

[19] A. Merchant, S. Batzner, S., S. S. Schoenholz, Scaling deep learning for materials discovery, Nature **624**, 80-85 (2023).

[20] A.G. Kusne, T. Gao, A. Mehta, L. Q. Ke, M. C. Nguyen, K.M. Ho, V. Antropov, C. Z. Wang, M. J. Kramer, C. Long, I. Takeuchi, On-the-fly machine-learning for high-throughput experiments: search for rare-earth-free permanent magnets. Sci. Rep. 4, 6367 (2014).

[21] A. Kabiraj, M. Kumar, S. Mahapatra, High-throughput discovery of high curie point two-dimensional ferromagnetic materials. npj Computational Materials 6, 35 (2020).

[22] J. Cai, X. Chu, K. Xu, H. Li, J. Wei, Machine learning-driven new material discovery. Nanoscale Advances 2, 3115 (2020).

[23] G. Katsikas, S. Charalampos, K. Joseph, Machine learning in magnetic materials. Phys. Stat. Sol. (b) 258, 2000600 (2021).

[24] T. D. Rhone, W. Chen, S. Desai, S. B. Torrisi, D. T. Larson, A. Yacoby, E. Kaxiras, Data-driven studies of magnetic two-dimensional materials. Sci. Rep. 10, 15795 (2020).

[25] G. A. Landrum, H. Genin, Journal of Solid State Chemistry 176. Application of machine-learning methods to solid-state chemistry: Ferromagnetism in transition metal alloys. J. Solid State Chem. 176, 587 (2003).

[26] C. W. Park, C. Wolverton, Developing an improved crystal graph convolutional neural network framework for accelerated materials discovery. Phys. Rev. Mater. 4, 063801 (2020).

[27] R. H. Wang, W. Y. Xia, T. J. Slade, X. Y. Fan, H. F. Dong, K. M. Ho, P. C. Canfield, C. Z. Wang, ML-guided discovery of ternary compounds involving La and immiscible Co and Pb elements. npj Computational Materials 8, 258 (2022).

[28] H. J. Sun, C. Zhang, W. Y. Xia, L. Tang, G. Akopov, R. H. Wang, K. M. Ho, K. Kovnir, C. Z. Wang, Machine learning guided discovery of ternary compounds containing La,

P and group IV elements. Inoganic Chem. 61, 16699 (2022).

[29] W. Xia, L. Tang, H.J. Sun, C. Zhang, K. M. Ho, G. Viswanathan, K. Kovnir, C. Z. Wang, Accelerating materials discovery using integrated deep machine learning approaches, J. Mater. Chem. A, 11 (47), 25973-25982 (2023).

[30] W. Xia, M. Sakurai, B. Balasubramanian, T. Liao, R Wang, C. Zhang, H. Sun, K-M. Ho, J. R. Chelikowsky, D. J. Sellmyer and C. Z. Wang, Accelerating the discovery of novel magnetic materials using a machine learning guided adaptive feedback, Proceedings of the National Academy of Sciences **119** (47), e2204485119 (2022).

[31] T. Xie, J. C. Grossman, Crystal Graph Convolutional Neural Networks for an Accurate and Interpretable Prediction of Material Properties, Phys. Rev. Lett. **120**, 145301 (2018).

[32] K. Choudhary, B. DeCost, Atomistic Line Graph Neural Network for improved materials property predictions, npj Comput Mater **7**, 185 (2021).

[33] M. Geiger, T. Smidt, e3nn: Euclidean Neural Networks, arXiv:2207.09453 (2022).

[34] G. Kresse and J. Furthmüller, Efficiency of ab-initio total energy calculations for metals and semiconductors using a plane-wave basis set, Comput. Mater. Sci. **6**, 15-50 (1996).

[35] G. Kresse and J. Furthmüller, Efficient iterative schemes for ab initio total-energy calculations using a plane-wave basis set. Phys. Rev. B **54**, 11169-11186 (1996).

[36] J.P. Perdew, K. Burke, K. and M. Ernzerhof, Generalized Gradient Approximation Made Simple, Phys. Rev. Lett. **77**, 3865-3868 (1996).

[37] P.E. Blöchl, Projector augmented-wave method. Phys. Rev. B **50**, 17953-17979 (1994).

[38] H.J. Monkhorst and J.D. Pack, Special points for Brillouin-zone integrations, Phys. Rev. B **13**, 5188-5192 (1976).

[39] S. Q. Wu, M. Ji, C.-Z. Wang, M. C. Nguyen, X. Zhao, K. Umemoto, R. M. Wentzcovitch, K.-M. Ho, An adaptive genetic algorithm for crystal structure prediction, J. Phys. : Condens. Matter **26**, 035402 (2013).

[40] X. Zhao, M. C. Nguyen, W. Y. Zhang, C. Z. Wang, M. J. Kramer, D. J. Sellmyer, X. Z. Li, F. Zhang, L. Q. Ke, V. P. Antropov, K. M. Ho, Exploring the Structural Complexity of Intermetallic Compounds by an Adaptive Genetic Algorithm, Phys. Rev. Lett. **112**, 045502 (2014).

[41] A. R. Oganov, C. W. Glass, Evolutionary crystal structure prediction as a tool in materials design, J. Phys. Cond. Matt. **20**, 064210 (2008).

**Table 1.** The compositions and structure symmetries of those Ce-Fe-X ternary compounds with $E_{hull}$ on the convex hull or within 50 meV/atom above the convex hull.

| | Phases | Symmetry | a (Å) | b (Å) | c (Å) | $E_{hull}$ (meV/atom) |
|---|---|---|---|---|---|---|
| **Existing stable phases [20]** | **$Ce_3FePb$** | *Pm-3m* | 5.11 | 5.11 | 5.11 | **0** |
| | **$Ce_3Fe_2In_2$** | *Pnma* | 16.98 | 4.83 | 7.52 | **0** |
| **ML+DFT predicted stable phases** | **$Ce_{12}Fe_7Bi$** | *I4/mcm* | 11.72 | 11.72 | 12.51 | **0** |
| | **$Ce_3FeBi$** | *P6₃/mmc* | 7.03 | 7.03 | 5.77 | **0** |
| | **$Ce_{12}Fe_7In$** | *I4/mcm* | 11.62 | 11.62 | 12.58 | **0** |
| | **$Ce_3Fe_2In_3$** | *C2/m* | 16.09 | 4.79 | 4.78 | **0** |
| | **$Ce_{12}Fe_3In_4$** | *C2/m* | 16.47 | 6.97 | 11.36 | **0** |
| | **$Ce_7Fe_4In$** | *Pm* | 6.84 | 5.04 | 7.46 | **0** |
| | **$Ce_{12}Fe_7Ag$** | *I4/mcm* | 11.52 | 11.52 | 12.64 | **0** |
| | **$Ce_6FeAg_{23}$** | *Fm-3m* | 13.57 | 13.57 | 13.57 | **0** |
| ML+DFT predicted metastable phases | $Ce_{12}FeBi_4$ | *Im-3m* | 9.72 | 9.72 | 9.72 | 4 |
| | $Ce_{15}FeBi_9$ | *P6₃mc* | 16.49 | 16.49 | 6.37 | 18 |
| | $Ce_{12}Fe_3Bi_4$ | *P2/m* | 5.80 | 6.97 | 12.23 | 32 |
| | $Ce_5FeBi_2$ | *Pnma* | 12.50 | 8.90 | 8.09 | 37 |
| | $Ce_{24}FeBi_8$ | *Pm-3m* | 9.39 | 9.39 | 9.39 | 40 |
| | $Ce_5FeBi_3$ | *P6₃/mcm* | 9.44 | 9.44 | 6.22 | 45 |
| | $Ce_{12}Fe_3Pb_4$ | *P2/m* | 5.79 | 6.91 | 12.19 | 23 |
| | $Ce_{12}Fe_6Pb$ | *Im-3* | 9.69 | 9.69 | 9.69 | 26 |
| | $Ce_5FePb_3$ | *P6₃/mcm* | 9.56 | 9.56 | 6.16 | 27 |
| | $Ce_{12}FePb_4$ | *Im-3m* | 9.63 | 9.63 | 9.63 | 46 |
| | $Ce_3Fe_2Pb_2$ | *Pbcm* | 5.96 | 8.79 | 12.74 | 47 |
| | $Ce_3FeIn$ | *Pm-3m* | 5.03 | 5.03 | 5.03 | 7 |
| | $Ce_2FeIn_2$ | *Pbam* | 5.03 | 5.03 | 5.03 | 11 |
| | $Ce_6FeIn_9$ | *I4/m* | 8.51 | 8.51 | 11.83 | 29 |
| | $Ce_{12}Fe_6In$ | *Im-3* | 9.65 | 9.65 | 9.65 | 29 |
| | $Ce_8FeIn_3$ | *P6₃mc* | 10.50 | 10.50 | 6.62 | 30 |
| | $Ce_{11}Fe_4In_9$ | *Cmmm* | 14.82 | 21.24 | 3.68 | 30 |
| | $Ce_4Fe_2In_3$ | *P2/m* | 7.66 | 3.66 | 8.07 | 34 |
| | $Ce_{12}Fe_6In$ | *Im-3* | 9.51 | 9.51 | 9.51 | 37 |
| | $Ce_8FeIn_{24}$ | *Pm-3m* | 9.36 | 9.36 | 9.36 | 38 |
| | $Ce_5FeIn_3$ | *P6₃/mcm* | 9.60 | 9.60 | 5.89 | 47 |
| | $Ce_7Fe_2In_3$ | *Cmmm* | 4.88 | 26.08 | 4.94 | 48 |
| | $Ce_{15}FeIn_9$ | *P6₃mc* | 16.39 | 16.39 | 6.08 | 49 |
| | $Ce_{11}Fe_4Ag_6$ | *I4/mmm* | 10.79 | 10.79 | 16.16 | 2 |
| | $Ce_{32}Fe_{19}Ag_3$ | *Amm2* | 9.28 | 15.87 | 15.66 | 2 |
| | $Ce_3FeAg$ | *Cmcm* | 6.99 | 12.00 | 5.54 | 19 |
| | $Ce_{12}Fe_6Ag$ | *Im-3* | 9.51 | 9.51 | 9.51 | 29 |
| | $Ce_3(FeAg)_2$ | *Pnma* | 16.57 | 4.86 | 7.12 | 32 |
| | $Ce_6FeAg_{32}$ | *Pm-3* | 9.24 | 9.24 | 9.24 | 33 |
| | $Ce_4Fe_2Ag_3$ | *P2/m* | 7.38 | 3.57 | 8.02 | 35 |
| | $Ce_{24}Fe_7Ag_8$ | *P2₁/c* | 8.80 | 12.03 | 18.21 | 35 |
| | $Ce_5Fe_2Ag$ | *I4/mcm* | 7.55 | 7.55 | 12.70 | 37 |
| | $Ce_9Fe_2Ag_3$ | *P2₁/c* | 6.64 | 11.95 | 16.97 | 37 |
| | $Ce_{24}Fe_7Ag_8$ | *P2₁/c* | 8.78 | 11.83 | 18.04 | 41 |
| | $Ce_{20}(Fe_3Ag)_3$ | *P422* | 7.49 | 7.49 | 12.83 | 46 |
| | $Ce_4FeAg$ | *F-43m* | 13.12 | 13.12 | 13.12 | 46 |
| | $Ce_3Fe_2Ag$ | *Pnma* | 9.97 | 4.77 | 10.61 | 47 |